\pdfoutput=1               
\documentclass[11pt]{article}

\usepackage[margin=1in]{geometry}
\usepackage{amsmath, amssymb, amsthm, proof, url, multirow}
\usepackage{booktabs} 
\usepackage{graphicx}
\usepackage{hyperref}
\usepackage{enumitem}
\usepackage{verbatim}
\usepackage{listings}
\usepackage{xspace}
\usepackage[T1]{fontenc}
\usepackage{lmodern}          
\usepackage{xcolor}
\usepackage[most]{tcolorbox}
\usepackage{enumitem}
\usepackage{microtype}
\usepackage{algorithm}          
\usepackage[noend]{algpseudocode} 
\algrenewcommand{\textproc}[1]{\textsf{#1}} 
\usepackage{xcolor} 
\lstdefinestyle{compacttt}{
  basicstyle=\ttfamily\scriptsize,   
  breaklines=true,
  breakatwhitespace=true,
  postbreak=\mbox{↪\space},          
  xleftmargin=2em,                   
  frame=single,
  aboveskip=0.5ex,
  belowskip=0.5ex
}

\lstdefinestyle{normaltt}{
  basicstyle=\ttfamily\normalsize,   
  breaklines=true,
  breakatwhitespace=true,
  postbreak=\mbox{↪\space},          
  xleftmargin=2em,                   
  frame=single,
  aboveskip=0.5ex,
  belowskip=0.5ex
}

\newenvironment{webexcerpt}
  {\begin{tcolorbox}[webcard]\sffamily\setlength{\parskip}{0.6em}\setlength{\parindent}{0pt}}
  {\end{tcolorbox}}

\newlist{webul}{itemize}{1}
\setlist[webul]{
  label=\textbullet,
  leftmargin=1.4em,
  itemsep=0.35em,
  topsep=0.35em,
  parsep=0pt
}

\newtheorem{definition}{Definition}
\newtheorem{theorem}{Theorem}[section]
\newtheorem{example}[theorem]{Example}
\theoremstyle{plain}            

\newcommand{\doctree}{D}
\newcommand{\GlobalCtx}[3]{\mathsf{GlobalCtx}_{#1}(#2,#3)}

\makeatletter
\@ifundefined{DfltCtxClosure}{\newcommand{\DfltCtxClosure}[1]{\mathsf{GlobalCtx}_{\mathsf{title}}\!\left(#1\right)}}
               {\renewcommand{\DfltCtxClosure}[2]{\mathsf{GlobalCtx}_{\mathsf{title}}\!\left(#1\right)}}
\@ifundefined{Keep}{}
                {}
\@ifundefined{Excerpt}{}
                  {}
\@ifundefined{Cl}{}
             {}
\@ifundefined{defeq}{\newcommand{\defeq}{\mathrel{\overset{\mathrm{def}}{=}}}}
                 {\renewcommand{\defeq}{\mathrel{\overset{\mathrm{def}}{=}}}}
\@ifundefined{wf}{}
              {}
\@ifundefined{serialize}{}
                      {}
\@ifundefined{treeify}{}
                    {}
\makeatother

\makeatletter
\@ifundefined{TitlePath}{\newcommand{\TitlePath}[1]{\mathsf{TitlePath}\!\left(#1\right)}}
                     {\renewcommand{\TitlePath}[1]{\mathsf{TitlePath}\!\left(#1\right)}}
\@ifundefined{HeadOne}{\newcommand{\HeadOne}[1]{\mathsf{H1}\!\left(#1\right)}}
                    {\renewcommand{\HeadOne}[1]{\mathsf{H1}\!\left(#1\right)}}
\makeatother

\makeatletter
\@ifundefined{Paths}{\newcommand{\Paths}[1]{\mathsf{Paths}\!\left(#1\right)}}
                 {\renewcommand{\Paths}[1]{\mathsf{Paths}\!\left(#1\right)}}
\@ifundefined{concat}{\newcommand{\concat}{\mathbin{\cdot}}}{}
\@ifundefined{Subtree}{}{}
\makeatother

\makeatletter
\@ifundefined{Cl}{}
             {}
\@ifundefined{restrict}{}
                    {}
\@ifundefined{BL}{}
               {}
\@ifundefined{Live}{}
                 {}
\@ifundefined{Excerpt}{}
                    {}
\makeatother

\makeatletter
\@ifundefined{Induced}{}
                    {}
\makeatother

\makeatletter
\@ifundefined{Render}{\newcommand{\Render}[2]{\mathsf{Render}\!\left(#1,#2\right)}}
                 {\renewcommand{\Render}[2]{\mathsf{Render}\!\left(#1,#2\right)}}
\@ifundefined{tok}{}
              {}
\@ifundefined{cost}{}
               {}
\@ifundefined{estcost}{}
                  {}
\makeatother

\newcommand{\Subdoc}[2]{\mathsf{Subdoc}(#1,#2)}
\newcommand{\Link}{\mathsf{Link}}

\makeatletter
\@ifundefined{Paths}{\newcommand{\Paths}[1]{\mathsf{Paths}(#1)}}{}
\@ifundefined{Subdoc}{\newcommand{\Subdoc}[2]{\mathsf{Subdoc}(#1,#2)}}{}
\@ifundefined{PrefixClosed}{}{}
\makeatother

\makeatletter
\@ifundefined{AdmissiblePrune}{}{}
\@ifundefined{MaxPaths}{}{}
\@ifundefined{Paths}{\newcommand{\Paths}[1]{\mathsf{Paths}(#1)}}{}
\@ifundefined{Subdoc}{\newcommand{\Subdoc}[2]{\mathsf{Subdoc}(#1,#2)}}{}
\@ifundefined{concat}{\newcommand{\concat}{\mathbin{\raisebox{0.2ex}{\tiny$\circ$}}}}{}
\@ifundefined{Subtree}{}{}
\makeatother

\makeatletter
\@ifundefined{ClosedView}{}{}
\@ifundefined{bllmfrac}{}{}
\makeatother

\title{SPIRE: Structure-Preserving Interpretable Retrieval of Evidence}
\author{
  Mike Rainey\\Carnegie Mellon University\\\texttt{mrainey@cmu.edu}
  \and
  Umut Acar\\Carnegie Mellon University\\\texttt{umut@cmu.edu}
  \and
  Muhammed Sezer\\Carnegie Mellon University\\\texttt{muhammedsezer12@gmail.com}
}
\date{\today}

\begin{document}

\maketitle
\begin{abstract}
Retrieval-augmented generation over semi-structured sources such as HTML is
constrained by a mismatch between document structure and the flat, sequence-based
interfaces of today’s embedding and generative models.  Retrieval pipelines
often linearize documents into fixed-size chunks before indexing, which
obscures section structure, lists, and tables, and makes it difficult to return
small, citation-ready evidence without losing the surrounding context that makes
it interpretable.

We present a structure-aware retrieval pipeline that operates over tree-structured
documents.  The core idea is to represent candidates as subdocuments:
precise, addressable selections that preserve structural identity while
deferring the choice of surrounding context.  We define a small set of document
primitives---paths and path sets, subdocument extraction by pruning, and two
contextualization mechanisms.  Global contextualization adds the non-local
scaffolding needed to make a selection intelligible (e.g., titles, headers, list
and table structure).  Local contextualization expands a seed selection within
its structural neighborhood to obtain a compact, context-rich view under a target
budget.  Building on these primitives, we describe an embedding-based candidate
generator that indexes sentence-seeded subdocuments and a query-time,
document-aware aggregation step that amortizes shared structural context.  We
then introduce a contextual filtering stage that re-scores retrieved candidates
using locally contextualized views.

Across experiments on HTML question-answering benchmarks, we find that preserving
structure while contextualizing selections yields higher-quality, more diverse
citations under fixed budgets than strong passage-based baselines, while
maintaining scalability.
\end{abstract}

\section{Introduction}

Retrieval-augmented generation (RAG) has become the dominant paradigm for
grounding language-model outputs in external evidence.
Yet most RAG pipelines begin by discarding the very structure that makes
documents intelligible: HTML pages, technical manuals, and reference articles
are flattened into sequences of fixed-size text chunks before indexing.
This linearization erases section boundaries, heading hierarchies, list and
table structure, hyperlinks, and other conventions that authors rely on to
organize meaning.
The result is a retrieval system that can find textually similar spans but
cannot reason about where those spans sit within a document, what structural
context they depend on, or how to present them as self-contained evidence.

The cost of this mismatch is most visible when retrieved evidence must be
shown to a user as a citation.
A sentence lifted from the middle of a numbered list, stripped of its list
heading and surrounding items, is often unintelligible.
A table cell without its row and column headers is meaningless.
A paragraph whose topic is established only by its section title becomes
ambiguous.
These are not edge cases; they are the norm for semi-structured web
documents, where meaning is distributed across the structural hierarchy
rather than concentrated in individual spans.

This paper introduces a retrieval framework that treats document structure as
a first-class object throughout the pipeline.
We parse HTML documents into tree-structured representations and assign every
node a stable, path-based identifier.
These paths serve simultaneously as citation-ready addresses and as the
primitive that retrieval algorithms use to select, merge, and expand
evidence.
Rather than committing to a flat rendering at indexing time, we work with
\emph{subdocuments}---path-addressed selections whose
contextualization is postponed until it is needed.

The framework is organized around two complementary forms of contextualization.
\emph{Global contextualization} enriches a retrieved node with the non-local
structural scaffolding required to make it interpretable in isolation: the
document title, enclosing section headers, list or table structure, and
similar cues that are logically associated with the node but may be distant
in the tree.
This global context is computed deterministically from the document structure
and prepended to each candidate at embedding time, producing embeddings that
encode both content and structural position.
\emph{Local contextualization} operates after retrieval, expanding a
sentence-level selection into its surrounding structural neighborhood.
An LLM-based filtering step then examines this expanded view to retain only
the material relevant to the query, yielding compact, context-rich excerpts.
Crucially, path identifiers are maintained through both stages, so that the
final citations point precisely to their source locations regardless of how
much context was added or removed during retrieval.

Figure~\ref{fig:sample-excerpts} illustrates the interaction between global
and local context.
The initial vector search returns sentence-level results that are rendered with
their global context.
Subsequent stages merge these results and selectively expand them to incorporate
local context, yielding excerpts that are both structurally grounded and
semantically complete.
This separation between global and local context allows retrieval algorithms to
reason explicitly about different sources of interpretability, rather than
conflating them through fixed-size chunking.

\tcbset{
  webcard/.style={
    enhanced,
    boxrule=0.4pt,
    colframe=black!20,
    colback=white,
    arc=2pt,
    outer arc=2pt,
    left=10pt,right=10pt,top=10pt,bottom=10pt,
    boxsep=0pt,
  }
}

\newcommand{\headertxt}[1]{\textit{\texttt{#1}}}

\begin{figure}
\centering
\resizebox{\linewidth}{!}{%
  \begin{minipage}{0.485\linewidth}
    \vspace{0pt}
    \begingroup\footnotesize

{\small\bfseries (1) First result sentence}\par
  \begin{webexcerpt}
    
    {\large\bfseries Explore The Natural Beauty Of State Parks In Virginia }\headertxt{[@H1]}\par

    Did you know \textbf{Virginia has over 41 state parks} bursting with \textbf{diverse landscapes }? \headertxt{[@H1,@S]}\par
  \end{webexcerpt}

  \vspace{0.5em}
{\small\bfseries (2) Second result sentence}\par
  \begin{webexcerpt}
    
    {\large\bfseries Explore The Natural Beauty Of State Parks In Virginia }\headertxt{[@H1]}\par

    {\normalsize\bfseries Key Takeaways} \headertxt{[@H1,@H2]}\par
    \begin{webul}
      \item Virginia has over 41 state parks with \textbf{diverse landscapes}, offering natural beauty and \textbf{outdoor activities} \headertxt{[@H1,@H2,@B1]}.
    \end{webul}
  \end{webexcerpt}

      \vspace{0.5em}
  {\small\bfseries (3) Merging of first two result sentences}\par
  \begin{webexcerpt}
    
    {\large\bfseries Explore The Natural Beauty Of State Parks In Virginia} \headertxt{[@H1]}\par

    Did you know \textbf{Virginia has over 41 state parks} bursting with \textbf{diverse landscapes}? \headertxt{[@H1,@S]}\par

    {\normalsize\bfseries Key Takeaways} \headertxt{[@H1,@H2]}\par
    \begin{webul}
      \item Virginia has over 41 state parks with \textbf{diverse landscapes}, offering natural beauty and \textbf{outdoor activities} \headertxt{[@H1,@H2,@B1]}.
    \end{webul}
  \end{webexcerpt}
    \endgroup
  \end{minipage}\hfill
  \hspace{0.5em}
  \begin{minipage}{0.485\linewidth}
    \vspace{0pt}
    \begingroup\footnotesize
    {\small\bfseries (4) Expansion of the merged result sentences}\par
      \begin{webexcerpt}
        
    {\large\bfseries Explore The Natural Beauty Of State Parks In Virginia} \headertxt{[@H1]}\par

    Did you know \textbf{Virginia has over 41 state parks} bursting with \textbf{diverse landscapes}? \headertxt{[@H1,@S1]}\par

    This blog will guide you through some of Virginia's most scenic state parks,
    highlighting their \textbf{unique attributes and attractions}. \headertxt{[@H1,@S2]}\par

    {\normalsize\bfseries Key Takeaways} \headertxt{[@H1,@H2]}\par
    \begin{webul}
      \item Virginia has over 41 state parks with \textbf{diverse landscapes}, offering natural beauty and \textbf{outdoor activities}. \headertxt{[@H1,@H2,@B1]}
      \item Some of the top state parks in Virginia include Grayson Highlands State Park, Shenandoah River State Park, Mason Neck State Park, Kiptopeke State Park, Pocahontas State Park, Natural Bridge State Park, and First Landing State Park. \headertxt{[@H1,@H2,@B2]}
      \item Scenic attractions within Virginia's state parks include Luray Caverns, Natural Tunnel State Park, Westmoreland State Park, Burke's Garden, the New River Trail State Park, Breaks Interstate Park, the Great Dismal Swamp, Sand Cave in Cumberland Gap National Historical Park, and Great Falls Park. \headertxt{[@H1,@H2,@B3]}
      \item Visitors can enjoy various \textbf{outdoor activities} at Virginia's state parks, such as hiking trails, camping, fishing, wildlife viewing, picnicking, and swimming. \headertxt{[@H1,@H2,@B4]}
    \end{webul}
  \end{webexcerpt}
  \endgroup
  \end{minipage}%
}
\caption{Sample excerpts generated while retrieving for the query:
``How many state parks are there in Virginia?''.
Items (1) and (2) show the initial results returned by vector search, where each
result corresponds to a single sentence.
In both cases, the sentence is rendered together with its \emph{global context},
which here consists of the document title and, for the second sentence, the
enclosing section header (``Key Takeaways'').
Item (3) illustrates the next stage of the pipeline, in which these two
sentence-level excerpts are merged into a single combined excerpt.
Finally, item (4) shows an expansion of the merged excerpt that supplies
additional \emph{local context}, incorporating content that is spatially adjacent
(above and below) to the merged region in the original document.
The labels shown in \headertxt{\{...\}} are meta-level annotations included only
for demonstration: each token denotes a structural element in the displayed
context (e.g., \headertxt{@H1} = document title, \headertxt{@H2} = section header,
\headertxt{@B1} = first bullet item), and these labels are not internal symbols of
the algorithm.
They are used to visualize what the contextualization rules include; for example,
\headertxt{[@H1,@H2,@B1]} in item (3) indicates that global contextualization of
bullet \headertxt{@B1} pulls in its enclosing header \headertxt{@H2} and title
\headertxt{@H1}.}

\label{fig:sample-excerpts}
\end{figure}

This two-stage design---global context for indexing, local context for
refinement---addresses a tension that runs through much of the RAG
literature.
Retrieving at a fine granularity (e.g., individual sentences) yields precise
hits but strips away context; retrieving at a coarse granularity (e.g., large
blocks) preserves context but dilutes relevance and wastes budget on
unhelpful material.
Our approach resolves this by decoupling the granularity of
\emph{matching}---which operates over globally contextualized
sentences---from the granularity of \emph{presentation}, which is determined
by local expansion and filtering under a token budget.

We situate this work within several threads of prior research.
Classical structured-document retrieval, notably the INEX
initiative~\cite{INEX2002}, established the problem of returning
sub-document elements at appropriate granularity from XML collections, but
predates neural embeddings and generative models.
Recent hierarchical retrieval systems such as
RAPTOR~\cite{sarthi2024raptor} build retrieval trees via clustering and
summarization, but construct the tree from content rather than exploiting the
document's own authored structure.
Similarly, Dense~X Retrieval~\cite{chen2024dense} shows that finer-grained 
retrieval units (propositions) outperform coarser passages. This result 
supports our choice of sentence-level retrieval, although their setting 
assumes flat documents rather than tree-structured ones. Other work also 
reports gains from finer-grained citation and retrieval 
units~\cite{zhang2024longcite,lin2025searchengine}.
Anthropic's contextual retrieval~\cite{anthropic2024contextual} enriches
chunks with LLM-generated context summaries before embedding---an approach
whose spirit our global contextualization shares, but which we implement
deterministically from document structure, making it cheaper, reproducible,
and provenance-preserving.
Late chunking~\cite{gunther2024late} takes a complementary approach,
encoding entire documents through a long-context transformer before chunking
the token embeddings so that each chunk inherits document-wide context via
attention.
Sentence-window and parent-document retrieval patterns decouple retrieval
granularity from synthesis granularity, but use flat offsets or fixed
parent--child mappings; our contextual filtering follows the document tree
and applies LLM-based filtering, with post-retrieval reranking following
the retrieve-then-rerank paradigm established by neural passage
reranking~\cite{nogueira2019passage}.
On the citation side, work on attributed
generation~\cite{bohnet2023attributed,gao2023enabling} and fine-grained
citation~\cite{zhang2024longcite,pmlr-v267-chuang25a} focuses on teaching
models to produce or verify citations, whereas our framework provides
structurally grounded citation units by construction.
Finally, work on context-window
utilization~\cite{liu2024lost,hong2025contextrot} shows that LLM performance
degrades as context length grows and when relevant information is buried in
the middle, motivating our approach of retrieving focused, structurally
contextualized fragments rather than large document blocks.

We make the following contributions:
\begin{enumerate}[leftmargin=*]
\item A \textbf{path-addressable document model} with a small set of
  primitives---paths, path sets, subdocument extraction, and two
  contextualization operators---that form a general interface between
  tree-structured documents and sequence-based models.

\item A \textbf{structure-aware retrieval pipeline} that indexes
  globally contextualized, sentence-seeded subdocuments; merges
  co-located results at query time; and applies local contextualization with
  LLM-based filtering to produce compact, citation-ready excerpts.

\item \textbf{Experimental evidence} on HTML question-answering benchmarks
  (HotpotQA and ASQA) showing that preserving structure yields
  higher-quality, more diverse citations under fixed token budgets compared
  to strong passage-based baselines, and that the contextual-filtering stage
  substantially improves citation precision.
\end{enumerate}

\section{Path-addressable document model}

Our goal for this section is to lift retrieval to a level of abstraction at which
retrieval systems can operate directly over
tree-structured documents.  This section focuses on the representation side of
that story: we introduce a small collection of primitives that let us name and
extract fine-grained evidence from an HTML-like tree while preserving enough
structure to keep the extracted result well formed and interpretable.  These
primitives serve as a common interface between document structure and the
sequence-based APIs of current models; the next section uses them to define the
retrieval algorithms themselves.

We proceed in four steps.  First, we define a path-addressable document model
in which every node in a document tree carries a stable, prefix-ordered
path.  Paths are citation-ready identifiers, and finite path sets
let us denote non-contiguous selections of nodes as a single unit of retrieval.
Second, we formalize a subdocument extraction operator that turns an
arbitrary path set into a well-formed subtree by completing the structural
context required to interpret the selection.  Third, we introduce global contextualization
policies, with a concrete policy for HTML, that enrich path sets in a
principled, idempotent way before extraction. Fourth, 
we introduce our approach to local contextualization, which expands a given
path set to include nearby content in the document.

\subsection{Document trees and paths}

We represent each source as a \emph{document} with explicit tree structure.
Formally, a document is a value of the inductive datatype $\mathbf{Doc}$:
\[
  \mathbf{Doc} \;::=\; \mathbf{Text}(s) \;\mid\; \mathbf{Elem}(\mathit{tag},\mathit{attrs},[\mathbf{Doc}]).
\]
Here $\mathbf{Text}(s)$ is a leaf carrying the string $s$, and
$\mathbf{Elem}(\mathit{tag},\mathit{attrs},[\mathbf{Doc}])$ is an interior node that carries an
HTML-like tag, an attribute map, and an ordered list of child documents.
Because this representation subsumes HTML, it provides a practical, uniform
encoding for a broad range of document types~\cite{crichton2024core,pandoc_3_8_3}.

To avoid ambiguity between a whole document and a particular node within it, we
treat nodes as positions in a document, identified by document paths.
A \emph{document path} is a finite list of child indices; we write path
concatenation as $p \concat [i]$.
Given a document $D$, we define $\Paths{D}$, the set of valid paths in $D$, and
$D[p]$, the (sub)document rooted at path $p$, inductively:
\begin{itemize}
  \item The root position has the empty path $[]$, and $[] \in \Paths{D}$ with $D[[]] = D$.
  \item If $p \in \Paths{D}$ and $D[p] = \mathbf{Elem}(\mathit{tag},\mathit{attrs},[D_1,\ldots,D_k])$,
        then for each $i \in \{1,\ldots,k\}$ we have $p \concat [i] \in \Paths{D}$ and
        $D[p \concat [i]] = D_i$.
\end{itemize}
Thus, when we refer to “a node in $D$”, we mean a path $p \in \Paths{D}$, and
$D[p]$ names the corresponding subtree.  These paths are fine-grained,
citation-ready identifiers for document content.

\newcommand{\Up}{\mathsf{Up}}
\newcommand{\Down}{\mathsf{Down}}
\newcommand{\Prune}[2]{\mathsf{Prune}(#1,#2)}
\makeatletter
\@ifundefined{PruneAux}{\newcommand{\PruneAux}[3]{\mathsf{PruneAux}(#1,#2,#3)}}{}
\makeatother

\begin{example}[Document tree and paths]
The HTML document in Listing~\ref{lst:tiny-html} parses to the document
tree shown in Listing~\ref{lst:tiny-parse-tree}.  Each subtree is
annotated with its path; for example, the text of the first paragraph
is identified by the path $[0,0,1,0]$.
\end{example}

\begin{lstlisting}[style=compacttt,frame=none,label={lst:tiny-html},caption={Tiny HTML document.}]
<html>
  <body>
    <section>
      <h1>Title</h1>
      <p>First paragraph.</p>
      <p>Second paragraph.</p>
    </section>
  </body>
</html>
\end{lstlisting}

\begin{lstlisting}[style=compacttt,frame=none,label={lst:tiny-parse-tree},caption={Document tree of the HTML document in Listing~\ref{lst:tiny-html}.}]
Elem("html", {}, [                   path=[]
  Elem("body", {}, [                 path=[0]
    Elem("section", {}, [            path=[0,0]
      Elem("h1", {}, [               path=[0,0,0]
        Text("Title",                path=[0,0,0,0])
      ]),
      Elem("p", {}, [                path=[0,0,1]
        Text("First paragraph.",     path=[0,0,1,0])
      ]),
      Elem("p", {}, [                path=[0,0,2]
        Text("Second paragraph.",    path=[0,0,2,0])
      ])
    ])
  ])
])
\end{lstlisting}

Whereas a single path identifies one node, retrieval and citation
typically involve multiple, non-contiguous regions of a document.
We therefore work with finite collections of paths.
A \emph{path set} is a finite subset $P \subseteq \Paths{\doctree}$.
Path sets allow us to refer to groups of document nodes, such as several
paragraphs, list items, or table cells, as a single unit of extraction,
retrieval, or citation.

Path sets give us a compact way to name multiple, possibly non-contiguous,
regions of a document.  However, an arbitrary path set does not by itself
guarantee that these regions can be interpreted meaningfully as part of a
document tree.
In practice, two kinds of structural completion arise repeatedly when working
with path sets:
\begin{itemize}
\item A selected node must be reachable from the root of the document.
If a path is included but one of its ancestors is missing, then the node cannot
be reached by descending the tree.
\item A selected structural node is often intended to stand for its
entire contents, not just the node boundary itself.
For example, selecting a paragraph typically means selecting its text as well.
\end{itemize}
We capture these two forms of completion using two simple path-set operations:
\begin{itemize}
  \item \emph{Ancestor completion.}
  Given a document tree \(\doctree\) and a path set \(P\), the operation
  \(\Up(\doctree,P)\) adds all ancestor paths required to reach every path in
  \(P\) from the root.
  Intuitively, \(\Up\) ensures connectivity.

  \item \emph{Descendant completion.}
  The operation \(\Down(\doctree,P)\) adds all descendant paths of each path in
  \(P\) that occur in \(\doctree\).
  Intuitively, \(\Down\) ensures that selecting a structural node includes its
  internal contents.
\end{itemize}
These two operations address complementary structural concerns, and we will
often apply them together.  We write
\[
  \Link(\doctree,P) \;\stackrel{\mathrm{def}}{=}\; \Up(\doctree,P)\ \cup\ \Down(\doctree,P)
\]
for this ancestor--descendant closure.

\begin{example}[Ancestor and descendant completion]
Let \(\doctree\) be the document tree in
Listing~\ref{lst:tiny-parse-tree}.

\emph{Ancestor completion.}
Suppose
\(P=\{[0,0,1]\}\), selecting the first paragraph node.
Then \(\Up(\doctree,P)\) adds the paths needed to reach this node from the root:
\[
\Up(\doctree,P)=\{[],[0],[0,0],[0,0,1]\}.
\]

\emph{Descendant completion.}
Applying descendant completion to the same set adds the paragraph’s text node:
\[
\Down(\doctree,P)=\{[0,0,1],[0,0,1,0]\}.
\]

\emph{Closure.}
Taking their union yields
\[
\Link(\doctree,P)=\{[],[0],[0,0],[0,0,1],[0,0,1,0]\},
\]
which recovers both the structural context and the internal content associated
with the selected node.
\end{example}

\subsection{Subdocuments}

Documents rarely need to be sent to a generative or an embedding model in their entirety.
Instead, we want to extract a small, well-formed subdocument from a source document
\(\doctree\) given an intensional description of relevant content as a set of paths
\(P \subseteq \Paths{\doctree}\).
A path set by itself is only a pointer-like representation (useful for naming and scoring);
to feed evidence to a model, or to render it for a reader, we must materialize it as
an actual document tree.

We do so in two steps.
First, we enrich \(P\) with the structural context needed for interpretation (at minimum via
ancestor--descendant closure \(\Link(\doctree,P)\), and later via global-contextualization policies).
Second, we extract the corresponding subdocument by deleting everything outside the enriched
set.  This second step is implemented by a pruning operator: it is not an alternative to path
sets, but the mechanism that turns a chosen set of paths into a concrete subtree.

\begin{definition}[Pruning operator]
Let $\doctree$ be a document tree and let $P \subseteq \Paths{\doctree}$ be a set of paths to keep.

We define an auxiliary pruning operator that prunes the subtree rooted at a path $r$:
\[
\PruneAux{\doctree}{r}{P} \;=\;
\begin{cases}
  \doctree'
    & \text{if } r \in P,\\
  \varepsilon
    & \text{otherwise},
\end{cases}
\]
where
\[
\doctree' \;=\;
\begin{cases}
  \mathbf{Text}(s)
    & \text{if } \doctree[r]=\mathbf{Text}(s),\\[4pt]
  \mathbf{Elem}(\mathit{tag},\mathit{attrs},\,\mathit{children})
    & \text{if } \doctree[r]=\mathbf{Elem}(\mathit{tag},\mathit{attrs},[\cdot]),
\end{cases}
\]
and
\[
\mathit{children}
\;=\;
\bigl[
  \PruneAux{\doctree}{r \concat [i]}{P}
\bigr]_{\,r \concat [i] \in P}.
\]
The bracketed child list discards any $\varepsilon$ results and preserves the original child order.

Finally, the (two-argument) pruning operator is defined by starting at the root:
\[
  \Prune{\doctree}{P} \;\stackrel{\mathrm{def}}{=}\; \PruneAux{\doctree}{[]}{P}.
\]
\end{definition}
\noindent
Intuitively, $\Prune{\doctree}{P}$ keeps exactly the nodes whose paths lie in $P$,
and removes everything else. In practice, $P$ must include enough ancestry
to reach the nodes of interest.

\begin{example}[Prune: keep two non-contiguous blocks (both paragraphs)]
Let $\doctree$ be the tree in Listing~\ref{lst:tiny-parse-tree}.  Consider the keep set
\[
P=\{[],[0],[0,0],
      [0,0,1],[0,0,1,0],
      [0,0,2],[0,0,2,0]\}.
\]
This keeps the two paragraph subtrees but drops the heading subtree at $[0,0,0]$.
The result of $\Prune{\doctree}{P}$ is shown in Listing~\ref{lst:prune1}.
\end{example}

\begin{lstlisting}[style=compacttt,frame=none,label={lst:prune1},caption={Both paragraphs only of our document in Listing~\ref{lst:tiny-parse-tree} (heading removed).}]
Elem("html", {}, [                   path=[]
  Elem("body", {}, [                 path=[0]
    Elem("section", {}, [            path=[0,0]
      Elem("p", {}, [                path=[0,0,1]
        Text("First paragraph.",     path=[0,0,1,0])
      ]),
      Elem("p", {}, [                path=[0,0,2]
        Text("Second paragraph.",    path=[0,0,2,0])
      ])
    ])
  ])
])
\end{lstlisting}

The example above already hints at what is awkward about using $\mathsf{Prune}$ directly:
to keep a node deep in the tree, the keep set must explicitly include all of its ancestors.
If we omit an ancestor, pruning cannot even reach the desired node.

\begin{example}[Prune: missing ancestry yields an unintended cut]
Let $\doctree$ be the tree in Listing~\ref{lst:tiny-parse-tree} and take
\[
P=\{[],[0],[0,0,1],[0,0,1,0]\},
\]
which tries to keep the first paragraph but forgets the \lstinline|section| node at path $[0,0]$.
Then $\Prune{\doctree}{P}$ cannot descend from $[0]$ to $[0,0,1]$, so the result collapses to
Listing~\ref{lst:prune2}.
\end{example}

\begin{lstlisting}[style=compacttt,frame=none,label={lst:prune2},caption={Result of pruning with a keep set that omits required ancestors.}]
Elem("html", {}, [                   path=[]
  Elem("body", {}, [                 path=[0]
  ])
])
\end{lstlisting}

Now, with prune and link, we define our subdocument operator.
Intuitively, a subdocument is the smallest well-formed document tree that
contains the selected nodes together with the structural context needed to
interpret them.  The definitions above make this intuition precise without
requiring upstream algorithms to reason directly about tree invariants.

\begin{definition}[Subdocument]
Given a document tree \(\doctree\) and a set of target paths
\(P \subseteq \Paths{\doctree}\), the \emph{subdocument} determined by \(P\) is
\[
  \Subdoc{\doctree}{P}
  \;\stackrel{\mathrm{def}}{=}\;
  \Prune{\doctree}{\Link(\doctree,P)}.
\]
\end{definition}

We illustrate the construction using the tiny HTML document and tree from
Listings~\ref{lst:tiny-html}–\ref{lst:tiny-parse-tree}.

\begin{example}[Subdocument: isolate a single paragraph]
Let \(P=\{[0,0,1]\}\), selecting the first paragraph node.

Ancestor completion adds the paths needed to reach this node from the root,
while descendant completion adds the paragraph’s text node.
Thus,
\[
\Link(\doctree,P)=\{[],[0],[0,0],[0,0,1],[0,0,1,0]\}.
\]
Pruning with this linked set yields the subdocument consisting of exactly the
first paragraph and its minimal context.
\end{example}

\begin{example}[Subdocument: non-contiguous blocks]
Suppose we wish to keep both the document title and the second paragraph.
Let
\(P=\{[0,0,0,0],\,[0,0,2,0]\}\).

Linking adds the necessary ancestors for both targets, producing a path set that
connects them through the document structure.
Pruning then yields a subdocument that contains these two blocks together,
even though they are not contiguous in the original document.
The result is shown in Listing~\ref{lst:subdoc2}.
\end{example}

\begin{lstlisting}[style=compacttt,frame=none,label={lst:subdoc2},caption={Subdocument: document title and second paragraph.}]
Elem("html", {}, [                   path=[]
  Elem("body", {}, [                 path=[0]
    Elem("section", {}, [            path=[0,0]
      Elem("h1", {}, [               path=[0,0,0]
        Text("Title",                path=[0,0,0,0])
      ])
      Elem("p", {}, [                path=[0,0,2]
        Text("Second paragraph.",    path=[0,0,2,0])
      ])
    ])
  ])
])
\end{lstlisting}

\makeatletter
\@ifundefined{DelayedSubdoc}{\newcommand{\DelayedSubdoc}[2]{\mathsf{Delayed}(#1,#2)}}{}
\@ifundefined{IsDelayedSubdoc}{\newcommand{\IsDelayedSubdoc}[2]{\mathsf{IsDelayed}(#1,#2)}}{}
\makeatother

In much of what follows, we will work not with a materialized subdocument
\(\Subdoc{\doctree}{P}\), but instead with the pair \((\doctree, P)\) itself.
At first glance, this may appear to be merely a trivial representation of a
path set, and it is worth explaining why this choice is both intentional and
important.

The key observation is that subdocument construction is typically
not something we want to perform eagerly.
Materializing \(\Subdoc{\doctree}{P}\) requires traversing and copying parts of
the document tree, an operation whose cost is proportional to the size of the
resulting tree.
In contrast, most upstream algorithms, such as global and local contextualization, operate by transforming or comparing
path sets, not by inspecting the concrete tree structure itself.

For this reason, we refer to the pair \((\doctree, P)\) as a \emph{subdocument}:
it denotes the intention to extract the subdocument induced by \(P\), without
actually performing the extraction yet.
All intermediate operations manipulate only the path set \(P\), while the
underlying document \(\doctree\) remains fixed.
The subdocument operator is applied only at the final stage, when we are ready
to render or serialize the result, for example, when constructing the input to
an embedding or a generative model.

Formally, the only invariant required of a subdocument is that its path
set be valid with respect to the document:
\[
  P \subseteq \Paths{\doctree}.
\]
This invariant is preserved by all of the transformations we consider.
As a result, subdocuments provide a uniform and lightweight interface
for reasoning about partial documents, while postponing the cost of tree
materialization until it is strictly necessary.

In the remainder of the paper, we therefore write \((\doctree, P)\) when referring
to a subdocument under construction, and reserve \(\Subdoc{\doctree}{P}\) for the
final, materialized tree that is passed to downstream consumers.

\subsection{Global contextualization}

Even when structurally sound, a subdocument may omit nearby contextual cues,
such as titles, section headers, list scaffolding, etc., that are important or even essential for interpretability
of its contents.  The
presence or absence of such cues substantially affects the quality of retrieval
and downstream reasoning.

To ensure that contextual information is reinstated when needed, we treat
context enrichment purely at the level of path sets.  Starting from any 
subdocument $(\doctree,P)$, a \emph{global contextualization} augments
\(P\) with additional paths mandated by a concrete policy~\(\theta\).  Applying
this enriched path set to the earlier subdocument operator yields a well-formed,
context-aware tree.

\begin{definition}[Global contextualization]
Let $(\doctree, P)$ be a subdocument.  
A \emph{global contextualization} for policy \(\theta\) is a function
\[
  \GlobalCtx{\theta}{\doctree}{\cdot} 
    : 2^{\Paths{\doctree}} \to 2^{\Paths{\doctree}}
\]
satisfying, for all subdocuments $(\doctree,P)$:
\begin{description}
  \item[\textsf{Extensivity}] \(P \subseteq \GlobalCtx{\theta}{\doctree}{P}\).
  \item[\textsf{Monotonicity}] If \(P \subseteq P'\), then 
        \(\GlobalCtx{\theta}{\doctree}{P}
          \subseteq \GlobalCtx{\theta}{\doctree}{P'}\).
  \item[\textsf{Idempotence}] \(
        \GlobalCtx{\theta}{\doctree}{\GlobalCtx{\theta}{\doctree}{P}}
        \;=\;
        \GlobalCtx{\theta}{\doctree}{P}.
        \)
\end{description}
\end{definition}
\noindent
Because \(\GlobalCtx{\theta}{\doctree}{P}\) is defined to be a subset of
\(\Paths{\doctree}\), it follows that global contextualization always preserves subdocuments.
Given a policy~\(\theta\), the context-enriched subdocument derived from
$(\doctree,P)$ is
\[
  \Subdoc{\doctree}{\GlobalCtx{\theta}{\doctree}{P}}.
\]

In the remainder of this section, we present the global-contextualization rule for HTML
titles in full detail, and then give summaries of the corresponding rules for
headers, lists, and tables.  
The complete HTML global-contextualization policy is obtained
by composing these individual rules.  This composition is well behaved: because
each rule is extensive, monotone, and idempotent, their composition preserves
these properties and therefore defines a valid global-contextualization operator.

\subsubsection{Titles}
Our first instance of global contextualization brings in the document title, wherever
it might be located in the document tree, either in a \lstinline|title| section
or, if there is none, in a header section.

\begin{definition}[Title global contextualization]
Let $\doctree$ be a document tree.
The \emph{title global contextualization} $\DfltCtxClosure{\doctree}$ is a fixed set of paths that
provides document‑wide context when no targets are selected.  We adopt the
HTML‑oriented policy:
\[
  \DfltCtxClosure{\doctree} \;\defeq\;
  \begin{cases}
    \{\TitlePath{\doctree}\},
      & \text{if a document title exists},\\[4pt]
    \{\HeadOne{\doctree}\},
      & \text{if a top‑level heading exists but no document title exists},\\[4pt]
    \{\},
      & \text{otherwise (empty context)}.
  \end{cases}
\]
Here $\TitlePath{\doctree}$ is the unique path to the document \texttt{<title>} node (if any),
and $\HeadOne{\doctree}$ is the path to the first \texttt{<h1>} node in document order (if any).
\end{definition}

\begin{example}[Title global context on the tiny tree]
On the tiny document from Listings~\ref{lst:tiny-html}–\ref{lst:tiny-parse-tree},
we typically have $\DfltCtxClosure{\doctree}=\{[0,0,0]\}$ (the \texttt{<h1>} at path $[0,0,0]$).
Thus:
\[
  \DfltCtxClosure{\doctree}=\{[0,0,0]\}.
\]
\end{example}

\subsubsection{Headers}

Beyond the document title, headers provide the most important form of global
context in structured documents. Unlike titles, which apply to the whole document, headers
scope their influence to the material that follows them. When a subdocument
selects a node, only the headers that are in effect at that point in the
document should be included in its context.

A complication is that, in HTML, header scoping is not determined purely by
tree ancestry. Headers such as \texttt{<h1>}–\texttt{<h6>} introduce regions of
influence that extend forward in document order and are terminated by later
headers of the same or higher rank. As a result, a paragraph may be governed by
a header that is not its ancestor in the DOM tree, but instead precedes it
linearly. Correctly recovering this structure requires reasoning about document
order and header levels rather than tree structure alone.

Our header global-contextualization rule captures this behavior by identifying, for each
selected path, the set of headers that are \emph{active} immediately before that
path in document order. Intuitively, these are the headers one would encounter
by scanning backward from the selected node, keeping the most recent header at
each level while discarding any that are superseded by intervening headers of
equal or higher rank. The global contextualization then adds exactly these active headers
to the path set.

This rule yields context that matches standard HTML reading semantics: content
under a section inherits its enclosing headers even when those headers are not
ancestors in the tree.

\makeatletter
\newcommand{\Tag}[2]{\mathsf{Tag}(#1,#2)}
\@ifundefined{Kids}{\newcommand{\Kids}{\mathsf{Kids}}}{}
\@ifundefined{Lists}{\newcommand{\Lists}{\mathsf{Lists}}}{}
\@ifundefined{Items}{\newcommand{\Items}{\mathsf{Items}}}{}
\@ifundefined{IsList}{\newcommand{\IsList}{\mathsf{IsList}}}{}
\@ifundefined{IsItem}{\newcommand{\IsItem}{\mathsf{IsItem}}}{}
\makeatother

\makeatletter
\@ifundefined{ItemAnc}{\newcommand{\ItemAnc}{\mathsf{ItemAnc}}}{}
\@ifundefined{ListAnc}{\newcommand{\ListAnc}{\mathsf{ListAnc}}}{}
\@ifundefined{ListSkel}{\newcommand{\ListSkel}{\mathsf{ListSkel}}}{}
\newcommand{\LinkM}{\mathsf{Link}_{M}}
\newcommand{\NearestItem}{\mathsf{NearestItem}}
\newcommand{\ListOf}{\mathsf{ListOf}}
\newcommand{\OnSpine}{\mathsf{OnSpine}}
\newcommand{\LeadKids}{\mathsf{LeadKids}}
\makeatother

\makeatletter
\@ifundefined{ChildOnPath}{\newcommand{\ChildOnPath}{\mathsf{ChildOnPath}}}{}
\@ifundefined{Spine}{\newcommand{\Spine}{\mathsf{Spine}}}{}
\@ifundefined{SpinePrefix}{\newcommand{\SpinePrefix}{\mathsf{SpinePrefix})}}{}
\makeatother

\subsubsection{Lists}

Lists introduce a subtle form of context because their semantic structure does
not align cleanly with simple subtree containment. In HTML, nested lists are
encoded by placing a sublist (\lstinline|<ul>| or \lstinline|<ol>|) inside a list
item (\lstinline|<li>|). As a result, a list item may simultaneously serve as a
label for a group (e.g.\ ``Lunch'') and as the structural parent of a nested
sublist (e.g.\ ``Sandwich'', ``Salad''). 
\[
\texttt{<ol>
  <li>Lunch
    <ul>
      <li>Sandwich</li>
      <li>Salad</li>
    </ul>
  </li>
</ol>}.
\]
Naively applying ancestor or descendant
closure in this setting either drops essential labels or pulls in unrelated
sibling items.

The key challenge is to include enough list structure to preserve meaning
without introducing extraneous siblings. When a subdocument selects an item
deep in a nested list, we typically want to retain (i) the item itself and its
immediate content, and (ii) the textual labels of enclosing list items that give
the item its hierarchical position. At the same time, we want to avoid including
other items from the same nested list unless they are explicitly selected.

Our list global-contextualization rule addresses this by treating list items asymmetrically.
For the innermost enclosing list item, the rule keeps the minimal path from that
item down to the selected node, preserving the item’s own content. For each
enclosing list item further out, the rule keeps only the leading children
that precede any nested sublist. This retains the label text (e.g.\ ``Lunch'')
while excluding the nested list itself, thereby preventing sibling items such as
``Sandwich'' from being pulled into the subdocument.

This construction yields context that matches standard reading semantics for
nested lists: selected items are accompanied by the labels that situate them,
but not by unrelated siblings.

\subsubsection{Tables}

Tables are tricky for global contextualization because a single cell is rarely meaningful
in isolation: its interpretation depends on the row and column labels that name
what the cell is “about.”  At the same time, pulling in an entire HTML table can
be extremely expensive in tokens, and much of the table (empty cells, unrelated
rows, decorative structure) is irrelevant when we are rendering or scoring cells
independently.

Our policy therefore treats table context as a lightweight labeling problem
rather than a reconstruction problem.  When a seed path set selects a table cell
(\lstinline|<td>| or \lstinline|<th>|), we augment the path set with just the
corresponding row label and column label cells, when they can be identified.
Concretely, we locate the enclosing table and row for the selected cell and
determine its column position within that row; we then add the leftmost header
cell in the row (row label) and the topmost header cell in the column (column
label).  This yields a minimal context that makes the selected cell intelligible
when rendered.

The key consequence is that the contextualization does not pull in other data cells or the
full table skeleton.  It adds only directly referenced labels, keeping the
result compact while preserving the associations that matter for downstream use
(e.g., embedding or generative-model scoring of individual cells).

\subsubsection{HTML-aware global contextualization}

\newcommand{\CtxClosureUnion}[2]{\GlobalCtx{\mathsf{HTML}}{#1}{#2}}
\newcommand{\CtxClosureTitle}[1]{\DfltCtxClosure{#1}}
\newcommand{\CtxClosureHdr}[2]{\GlobalCtx{\mathsf{hdr}}{#1}{#2}}
\newcommand{\CtxClosureLst}[2]{\GlobalCtx{\mathsf{lst}}{#1}{#2}}
\newcommand{\CtxClosureTbl}[2]{\GlobalCtx{\mathsf{tbl}}{#1}{#2}}

Thus far we have specified global-contextualization rules targeted at particular HTML
structures, that is, titles, headers, lists, and tables.  These rules 
capture the minimal dependencies needed for a
fragment to be intelligible.  However, our downstream uses of global contextualization
typically require a single, uniform notion of global contextualization that we can
apply to any seed set of paths.  As a first, deliberately simple step,
we define a composite rule that merely takes the union of the four specialized
policies.  

We refer to this rule as the \emph{HTML-aware global contextualization} and write it
\(\CtxClosureUnion{N}{P}\).

\begin{definition}[HTML-aware global contextualization]
\label{def:union-global-contextualization}
Let $(\doctree,P)$ be a subdocument.
Write
\[
\begin{aligned}
  \CtxClosureUnion{N}{P}
  \;\stackrel{\mathrm{def}}{=}\;
  P
  &\;\cup\;
  \CtxClosureTitle{N} \\[2pt]
  &\;\cup\;
  \CtxClosureHdr{N}{P}
  \;\cup\;
  \CtxClosureLst{N}{P}
  \;\cup\;
  \CtxClosureTbl{N}{P}.
\end{aligned}
\]
\end{definition}

The operator simply aggregates whatever additional paths each specialized rule
would introduce: (i) title-level context via \(\CtxClosureTitle{N}\) (as
specified in the title subsection), (ii) enclosing-section scaffolding via
\(\CtxClosureHdr{N}{P}\), (iii) list spines via \(\CtxClosureLst{N}{P}\), and
(iv) table labels via \(\CtxClosureTbl{N}{P}\).
Because each constituent is extensive, monotone, and idempotent on its domain,
their union is again a global-contextualization operator.

\subsubsection{Potential for extensibility}

We illustrated a policy for global contextualization drawn from familiar HTML structure. 
The larger point is that the mechanism is extensible: a contextualization policy can be specialized to 
other document idioms without changing the subdocument pipeline. One common and feasible 
specialization is worth highlighting.
Many technical and legal documents maintain a dedicated section where terms are defined 
(e.g., a glossary or “Definitions”). In HTML this is routinely encoded with stable cues: 
anchors/IDs on definition blocks, \lstinline|<dfn>| or \lstinline|<dt>/<dd>| pairs, or lightweight data-attributes. A glossary-aware policy can:
(i) pre-index definitions by canonical term (and synonyms, if annotated);
(ii) recognize term mentions in the body via explicit links, \lstinline|<dfn>| references, or scoped annotations; and
(iii) add the corresponding definition block (and optionally a small, fixed neighborhood as scaffolding) 
to the context whenever a mention appears in $P$.
Because this policy only adds well-identified paths, it preserves the basic algebraic properties of global contextualization
(extensive, monotone, idempotent with memoization) and integrates cleanly with masked pruning to include definition 
shells without dragging in unrelated content.

\subsection{Local contextualization}
Global contextualization enriches a selected path set with non-local structural
scaffolding (e.g., titles, active headers, list spines, and table labels) so that
an extracted view remains interpretable.
However, even an interpretable subdocument can be too small to serve as
a self-contained unit of evidence for downstream components.
For example, a sentence-level selection is often precise but underspecified
without nearby sentences or the surrounding paragraph.

We therefore introduce a complementary mechanism that expands a 
subdocument using only local structural neighborhoods in the document tree,
with the goal of producing a compact but context-rich view under a target budget
(e.g., a token budget for a chosen model).
We refer to this process as \emph{local contextualization}.
Given a subdocument $(\doctree,S)$, the goal is to derive a transient
expanded view $(\doctree,P)$ that:
\begin{enumerate}
  \item includes $S$;
  \item approaches a target input size (e.g., in terms of words, tokens, etc.);
  \item respects document structure, favoring complete sentences, paragraphs,
        and block-level elements; and
  \item preserves path stability, so that results from different expansions can
        be compared and combined.
\end{enumerate}
Local contextualization and global contextualization are orthogonal and can be
composed: the former grows a selection outward in place, while the latter adds
structural cues that may be essential for interpretation.

Local contextualization proceeds by gradually enlarging the path set $S$ using structural
relations in the document tree.
At each step, we consider adding surrounding context drawn from:
\begin{itemize}
  \item sentence completion, to avoid dangling fragments;
  \item enclosing block-level elements (e.g., paragraphs, list items); and
  \item nearby regions in document order.
\end{itemize}
Each candidate expansion is evaluated by rendering the resulting subdocument and
estimating its size under the chosen size function (e.g., a tokenizer).
Expansion continues until the rendered view is close to the target size, or no
further admissible growth is possible without exceeding it.

\paragraph{Contrast with fixed-size chunking.}
Many retrieval pipelines rely on fixed-size chunkers that partition documents
into overlapping spans of tokens or characters prior to indexing.
Such chunkers are agnostic to document structure: they may split paragraphs,
interleave unrelated sections, or duplicate content across overlapping windows.
In contrast, our expansion mechanism operates after an initial selection, starts
from a precise subdocument, and grows outward in a structure-aware
fashion, prioritizing complete sentences, paragraphs, and block-level elements.
This lets us tailor the presented context to a target budget without committing
to a single, global chunking of the corpus.

\newcommand{\CtxClosureCost}[3]{\mathsf{CtxClosureCost}_{#1}(#2,#3)}
\newcommand{\MergeStep}{\mathsf{MergeStep}}
\newcommand{\SentPaths}{\mathsf{SentPaths}}
\makeatletter
\@ifundefined{InorderSeq}{\newcommand{\InorderSeq}[1]{\mathsf{IOSeq}\!\left(#1\right)}}{}
\@ifundefined{NextIO}{\newcommand{\NextIO}[2]{\mathsf{Next}_{\mathsf{io}}\!\left(#1,#2\right)}}{}
\@ifundefined{PrevIO}{\newcommand{\PrevIO}[2]{\mathsf{Prev}_{\mathsf{io}}\!\left(#1,#2\right)}}{}
\@ifundefined{NeighborIO}{\newcommand{\NeighborIO}[2]{\mathsf{Nbr}_{\mathsf{io}}\!\left(#1,#2\right)}}{}
\@ifundefined{ZoomNeighbors}{\newcommand{\ZoomNeighbors}[2]{\mathsf{ZoomNbrs}\!\left(#1,#2\right)}}{}
\@ifundefined{ZoomStep}{\newcommand{\ZoomStep}[2]{\mathsf{ZoomStep}\!\left(#1,#2\right)}}{}
\makeatother

\newcommand{\Sub}{\mathsf{Sub}}
\newcommand{\doccorpus}{\mathcal{D}}
\newcommand{\acitation}{c}
\newcommand{\citations}{C}
\newcommand{\userquery}{Q}
\newcommand{\Rel}{\mathsf{Rel}}
\newcommand{\tokcount}[1]{b_{#1}}
\newcommand{\llmmodel}{M}
\newcommand{\Cost}{\mathsf{Cost}}

\section{Structure-aware retrieval pipeline}
\label{sec:retrieval}

This section connects the document model developed in the previous section
with retrieval techniques familiar from the literature on RAG.
Our aim is twofold.
First, for readers experienced with RAG, vector search, or hybrid retrieval
pipelines, we show how our notion of subdocuments fits naturally
into conventional embedding- and generative-model-based retrieval workflows.
Second, for readers with a programming-languages background, we explain how
ideas such as explicit structure, delayed materialization, and contextualization
operators provide a principled foundation for retrieval over semi-structured
documents such as HTML.

At a high level, our retrieval pipeline has two stages.
An \emph{embedding-based} stage performs high-recall similarity search over a
fixed inventory of candidates.
An optional \emph{contextual filtering} stage then reasons over the retrieved candidates to
refine or validate relevance.
Although these stages can be used independently, we present them together to
emphasize the role of subdocuments as a unifying representation.

\paragraph{Subdocuments as retrieval units.}
In our system, the natural unit of retrieval is the subdocument.
This contrasts with many RAG pipelines, where the retrieval unit is typically a
text span of some kind (e.g., byte/character offsets, line, column ranges,
or other start/end indices into a flattened text view), or a pre-flattened
hierarchy such as sections containing paragraphs containing
sentences represented as strings.
By using subdocuments, we preserve a precise connection to the original
document structure, including semantically meaningful markup such as links and
formatting, and we can represent retrieval units at many granularities (a
sentence, a paragraph, a table cell, etc.) within a single uniform datatype.

We apply global contextualization and rendering only when required by downstream consumers, rather than treating the contextualized
excerpt as the retrieval unit.
Doing so would prematurely discard potentially useful structure: once a seed
selection (say, a sentence or HTML element) is replaced by its global contextualization,
we lose the ability to distinguish what was originally selected from what was
added for intelligibility.
Subdocuments therefore let retrieval operate over precise, structured
selections while postponing contextual augmentation to the point where it is
actually needed.

\begin{figure}
\centering

\begin{minipage}{0.48\linewidth}
\begin{lstlisting}[style=compacttt,frame=none]
<section>
  <h2>Background</h2>
  <p>
    <sentence-beg/>
    Ada Lovelace wrote
    the first algorithm.
    <sentence-end/>
    <sentence-beg/>
    <a href="...">Her notes</a>
    described the Analytical Engine.
    <sentence-end/>
    </a>
  </p>
</section>
\end{lstlisting}
\end{minipage}
\hfill
\begin{minipage}{0.48\linewidth}
\[
\begin{array}{l}
\textbf{Sentence seed:} \\[0.3ex]
\quad S_b = \SentPaths(\doctree,b)
\\[1ex]
\textbf{Subdocument:} \\[0.3ex]
\quad (\doctree, S_b)
\\[1ex]
\textbf{Global contextualization:} \\[0.3ex]
\quad P_b = \GlobalCtx{\mathsf{HTML}}{\doctree}{S_b}
\\[1ex]
\textbf{Rendered for embedding:} \\[0.3ex]
\quad \Render{\doctree}{P_b}
\end{array}
\]
\end{minipage}

\vspace{0.75ex}

\caption{%
Sentence-anchored subdocuments.
Sentence boundaries are inferred and recorded using zero-width markers without
altering the HTML tree.
Each sentence induces a seed path set $S_b$, yielding a subdocument
$(\doctree,S_b)$ that is stored in the embedding index 
(not contextualized).
Global contextualization and rendering are applied only at embedding time, producing a
self-contained excerpt suitable for similarity search while preserving a precise
reference back into the original document structure.
}
\label{fig:sentence-anchored-subdoc}
\end{figure}

\subsection{Vector-embedding-based retrieval}
\label{sec:vector-retrieval}

We begin with vector-embedding-based retrieval, which provides an efficient and
query-agnostic mechanism for candidate generation.
As in standard RAG pipelines, we embed a fixed collection of units, embed the
user query, and retrieve the most similar units.
A key design choice is the unit of indexing.
We use sentences as the embedding seeds: empirically, sentence-level units strike
a good balance between precision and recall, while paragraphs are often too large
and semantically mixed, which can dilute similarity signals and degrade
retrieval quality.
Broader context is added later by contextualization and expansion, rather than
being baked into the indexed unit.

\paragraph{Indexing surface.}
Let $\doccorpus = \{\doctree_1,\ldots,\doctree_m\}$ be a corpus of document trees.
An \emph{indexing surface} is a finite set
\[
  \mathcal{I} \subseteq
  \bigl\{ (\doctree,P) \mid \doctree \in \doccorpus,\ P \subseteq \Paths{\doctree} \bigr\},
\]
whose elements are subdocuments.
In this section we assume that, for each document tree $\doctree$, we can
enumerate its sentences as path sets.
Concretely, each sentence is identified by a begin path $b \in \Paths{\doctree}$
and a corresponding sentence seed path set $S_b \subseteq \Paths{\doctree}$ that
selects exactly that sentence.
Our embedding index stores the sentence-seeded subdocuments
$(\doctree,S_b)$.

\paragraph{From subdocuments to embeddings.}
Figure~\ref{fig:sentence-anchored-subdoc} illustrates our pipeline.
Given a subdocument $(\doctree,P)$ (in our case, $P=S_b$), we:
\begin{enumerate}
  \item apply HTML-aware global contextualization to obtain an enriched path set $P'$;
  \item materialize the subdocument $(\doctree, P')$;
  \item render the result to a textual format of choice (our implementation uses Markdown); and
  \item embed the rendered text.
\end{enumerate}
Thus, the indexed unit and the embedded text are intentionally distinct: the
former preserves precision and referential stability, while the latter provides
enough context for semantic similarity.

\subsubsection{Supporting sentence-level references}

Unlike many RAG settings, our input documents do not come with sentence
structure built in.
HTML provides elements such as paragraphs, list items, and table cells, but
these are often too coarse: a single paragraph may contain multiple unrelated
sentences, while splitting paragraphs na\"ively risks destroying inline
structure such as hyperlinks or emphasized text.
Our solution is to derive sentence structure while preserving the original
HTML tree.
As a preprocessing step, we apply an off-the-shelf sentence segmenter to the
rendered text of each block-level leaf.
The resulting character spans are then mapped back to the HTML tree, where we
insert zero-width sentinel elements
\[
  \text{\lstinline|<sentence-beg/>|} \quad\text{and}\quad
  \text{\lstinline|<sentence-end/>|}
\]
at the corresponding boundaries.
These sentinels do not reparent or duplicate content; instead, they annotate the
existing structure with explicit sentence boundaries while preserving all
semantic markup, including links and inline formatting.
Each sentence is thereby associated with a distinguished begin-sentinel path
$b$, and the set of paths belonging to that sentence can be recovered as
\[
  \SentPaths(N,b)
  \;\stackrel{\mathrm{def}}{=}\;
  \{\, p \in \Paths{N} \mid p \text{ is annotated with } b \,\}.
\]

\subsubsection{Query-time retrieval}
At query time, the user query is embedded and compared against the vectors in the
index, yielding an ordered list of subdocuments ranked by similarity.
In our experiments (and in common RAG evaluations), results are presented under a
fixed output budget (e.g., a token budget used to match passage-based baselines and
to reflect downstream input limits).
A naive strategy is therefore to take the top-$k$ subdocuments (or to admit
items greedily until the budget is exhausted) after applying global contextualization
to each item independently.

In our setting, this naive strategy performs poorly because of an interaction between
(1) precise, overlapping indexed units and (2) document-level global contextualization.
Sentence-seeded subdocuments are intentionally fine-grained, so many highly
ranked items can originate from the same document and share substantial structural
scaffolding (document title, enclosing headers, list structure, etc.).
If we global-contextualize and render each retrieved item in isolation, this shared
scaffolding is repeatedly included and repeatedly charged against the budget.
As a result, near-duplicate outputs from a single document can crowd out other relevant
evidence and reduce effective recall across documents.

A typical failure mode looks like the following: the ranker returns two items from the
same document early in the list, both of which global-contextualize to include the
same title and first header. The earlier item contributes little beyond this shared
context, while the later item contains the answer-bearing sentence under that header.
Costing them separately can cause the shared title/header material to consume enough of
the budget that the answer-bearing item is excluded, even though it was retrieved.

\begin{example}[Redundancy under global contextualization]
Consider a document \(\doctree\) with title \textsf{Foo} and a first section header
\textsf{First header of Foo}. Suppose the similarity ranker returns two sentence-seeded
subdocuments from \(\doctree\), \(x_1=(\doctree,S_{b_1})\) and \(x_2=(\doctree,S_{b_2})\),
with \(x_1\) ranked above \(x_2\).
After global contextualization and rendering, both items include the same structural
scaffolding:
\[
\begin{aligned}
\mathsf{Render}\!\left(
  \Subdoc{\doctree}{\GlobalCtx{\mathsf{HTML}}{\doctree}{S_{b_i}}}
\right)
&\approx
\texttt{\# Foo}\;+\;\texttt{\#\# First header of Foo} \\
&\qquad+\;\text{(local content near }b_i\text{)}.
\end{aligned}
\]
If \(x_1\) contains only a generic sentence under the header, while \(x_2\) contains the
answer-bearing sentence, then charging each rendered item independently can consume the
budget on two near-duplicates whose shared title/header dominates their cost.
In that case, \(x_2\) may be excluded even though it was retrieved and ranked highly.
Aggregating by document avoids this outcome: we union the path sets
\(S_{b_1}\cup S_{b_2}\), global-contextualize once for \(\doctree\), and render a single view
that pays for the title and header once while retaining both sentences.
\end{example}

\paragraph{Document-aware aggregation.}
To avoid repeatedly paying for shared structure, we aggregate retrieved items at the
document level before enforcing the final budget.
Concretely, we:
\begin{enumerate}
  \item retrieve an initial prefix of the similarity-ranked subdocuments;
  \item group candidates by their source document tree $\doctree$;
  \item for each document, merge candidate path sets by union to form one aggregated
        subdocument $(\doctree, P_{\doctree})$;
  \item apply global contextualization once per document to each merged selection and
        compute its rendered cost; and
  \item select the highest-ranked aggregated results whose combined cost fits the target budget.
\end{enumerate}
If the selected, deduplicated set falls well below the target budget, we increase the
retrieval-prefix bound and repeat, admitting additional candidates until the aggregated
result set approaches the budget.

\subsubsection{Summary}
The output of query-time retrieval is a sequence of subdocuments, with at most one
aggregated result per source document, whose combined post-contextualization cost respects
the target budget while preserving the similarity ranking as closely as possible.
This document-aware aggregation is crucial for retrieval over  
subdocuments: it amortizes shared structural context (titles, headers, scaffolding) that
would otherwise be duplicated across many fine-grained hits from the same document, and it
keeps embedding retrieval competitive with conventional passage-based baselines without
sacrificing referential precision.

Our choice of sentences as the unit of embedding is not essential to our overall
approach of using path-addressable subdocuments.
Other units are compatible with our approach, including ones using mixed granularities,
e.g., mixing sentence- and paragraph-level embeddings.
The only requirement is that the unit be representable by a path set.

\subsection{Contextual filtering}
\label{sec:generative-retrieval}

The embedding stage performs high-recall candidate generation over the full corpus
$\doccorpus$, but it does so using small indexing units (in our case, sentence-seeded
subdocuments).  This granularity is useful for retrieval---it yields precise,
citation-ready anchors---but it is often too small to support reliable downstream
decisions.  In particular, the relevance of a sentence frequently depends on nearby
content (adjacent sentences, the enclosing paragraph, or a local list/table region),
and this information is intentionally not part of the indexed unit.

For this reason, we introduce a second, query-time filtering stage that evaluates
retrieved candidates in their local structural neighborhood.
This stage does not search the corpus anew; it operates on a finite set of candidates
produced by embedding retrieval, which we call a subdocument surface:
\[
\mathcal{S} \subseteq
\bigl\{(\doctree,P) \mid \doctree \in \doccorpus,\ P \subseteq \Paths{\doctree}\bigr\}.
\]
Each element of $\mathcal{S}$ is a subdocument that precisely identifies
a region of interest (often anchored at a sentence), but does not yet commit to a
particular amount of surrounding context.

\paragraph{Why filtering needs local context.}
Embedding similarity can conflate fine-grained distinctions (dimension leakage), and
sentence-level units are prone to ambiguity when viewed in isolation (references,
scope, negation, and exceptions often sit just outside the sentence boundary).
As a result, a candidate that appears highly similar at the sentence level may become
clearly irrelevant once nearby context is included, and conversely, a candidate may only
become interpretable when expanded to include its immediate neighborhood.

\paragraph{Filtering via local contextualization.}
The filtering stage therefore evaluates each candidate after expanding it with local
context.
Concretely, for each $(\doctree,P)\in\mathcal{S}$ we derive an expanded view by applying
local contextualization (structure-aware expansion) to obtain a larger path set $P^{+}$
that still contains $P$ but includes nearby structural context.
We then pass to a query-conditioned scoring function the expanded view, and have the function decide
which candidates to retain and which to discard.

\subsubsection{Scoring citations via generative language models}
In our implementation this scoring function is implemented by a generative model, prompted with the query and the
expanded view.
After structure-aware expansion, the generative model is applied to each expanded
view derived from a subdocument in $\mathcal{S}$.
In each invocation, we present the model with the user query $Q$ together with a
single expanded view $(\doctree,P)$, rendered to text.
The task of the model is not to produce an answer directly, but to identify which
parts of the view are essential for answering $Q$.

To make this task precise and robust, we expose the internal citation structure
of the view explicitly.
Before rendering, we identify a set of \emph{citable units} within $P$, such as
sentences, list items, or paragraphs, each corresponding to a path-stable region
of the document.
When rendering the expanded view, we wrap each such unit $i$ with a lightweight,
deterministic marker
\lstinline|<lab_i>| $\;\cdots\;$ \lstinline|</lab_i>|
inserted on the fly, without modifying the underlying document tree.

The generative model is then instructed to return exactly the set of labels whose
corresponding units are required to answer $Q$.
Each label $\ell$ deterministically maps back to a path set
$P(\ell) \subseteq \Paths{\doctree}$, yielding a citation
$(\doctree, P(\ell))$.
Aggregating over all expanded views, we obtain the final citation set
\[
  C^{\star} 
  \;\triangleq\; 
  \bigcup_{\text{expanded views}}
  \{\,(\doctree,\,P(\ell)) \mid \ell \text{ selected by the model} \,\}.
\]

\paragraph{Example: generative-model prompt from a subdocument.}
We illustrate how the running example from
Figure~\ref{fig:sentence-anchored-subdoc} is presented to a generative model.
Suppose the embedding-based stage retrieves the sentence-anchored 
subdocument $(\doctree,S_b)$ corresponding to the sentence
\emph{``Her notes described the Analytical Engine.''}
Structure-aware expansion then produces a context-rich view $(\doctree,P)$,
including the enclosing paragraph and section header.

Before rendering, we identify sentence- or paragraph-level citable units within
$P$ and assign each a deterministic label.
The resulting prompt presented to the generative model has the following form:
\begin{lstlisting}[style=compacttt,frame=none]
User query:
Who described the Analytical Engine?

Document excerpt:
## Background

<lab_1>Ada Lovelace wrote the first algorithm.</lab_1><lab_2>Her notes described
the Analytical Engine.</lab_2>

Instruction:
Select the labels of all excerpts that are required to answer the query.
Return only a JSON list of labels.
\end{lstlisting}

In this example, the generative model is expected to return
\lstinline|["lab_2"]|.
The label \lstinline|lab_2| deterministically maps back to a context-closed path
set $P(\texttt{lab\_2}) \subseteq \Paths{\doctree}$, yielding a citation
$(\doctree,P(\texttt{lab\_2}))$.
Crucially, although the model reasons over an expanded, rendered view, the
citation is recorded in terms of the original subdocument structure.

This label-based protocol ensures that citation extraction is stable under
re-rendering and robust to paraphrasing or normalization performed by the
generative model.
In contrast to schemes that ask the model to reproduce citation text verbatim and
recover citations by string matching, our approach preserves a precise link to
the original document structure.
The generative model reasons over expanded, context-rich views, while citations
are recorded as subdocuments with well-defined structural identity.


\newcommand{\sysBGE}{\textsc{bge}\xspace}
\newcommand{\sysEmbed}{\textsc{embedding\_retrieval}\xspace}
\newcommand{\sysEmbedNoCtx}{\textsc{embedding\_retrieval\_noctx}\xspace}
\newcommand{\sysGenerative}{\textsc{SPIRE}\xspace}
\newcommand{\datasetHotpot}{\textsc{HotpotQA}\xspace}
\newcommand{\datasetASQA}{\textsc{ASQA}\xspace}

\section{Experimental evidence}
\label{sec:evaluation}

This section evaluates the retrieval algorithms proposed in the paper.
Our experiments are organized around two questions.
First, we ask whether our vector-embedding-based retrieval algorithm is competitive
with a strong HTML-specific baseline, \sysBGE, when both systems use the same
embedding model and are evaluated under a fixed citation budget.
Second, we evaluate whether augmenting embedding-based retrieval with a
contextualized-filtering stage improves citation quality.
Across both sets of experiments, we measure retrieval quality using an LLM-as-judge
protocol that assesses whether individual citations are helpful for answering the
query.
Together, these experiments isolate the effects of citation granularity,
global contextualization, and contextual filtering, and demonstrate how each
contributes to producing higher-quality, budget-efficient citation sets.

\subsection{Experimental setup}

\paragraph{Rubric.}
We evaluate citation quality using an LLM-as-judge procedure.
For each query--citation pair, we prompt a judge model to determine whether the
citation is \emph{helpful} for answering the query (binary decision).
This evaluation style follows prior work; in particular, SelfCite~\cite{pmlr-v267-chuang25a}
uses an LLM judge to assess citation quality.
As in their setting, we validated the judge by manually inspecting a random sample of
judgments for consistency and obvious failure modes.
For all experiments reported here, the judge is \emph{Qwen~3 235B Instruct}~\cite{qwen3-235b-a22b-instruct-modelcard}, and the prompt
we use appears in Appendix~\ref{sec:prompts}.
We report results as \(\#\mathrm{helpful}/\#\mathrm{total}\) and the corresponding ratio.

\paragraph{Datasets.}
We evaluate on two QA benchmarks used in HtmlRAG: \datasetHotpot{}~\cite{yang-etal-2018-hotpotqa}
and \datasetASQA{}~\cite{stelmakh-etal-2022-asqa}.
Following HtmlRAG~\cite{10.1145/3696410.3714546}, we use their released evaluation suites:
for each benchmark, HtmlRAG samples 400 instances uniformly at random and augments each instance
with HTML pages sourced via Bing search. We use this data as-is for all reported runs, so
our results are directly comparable to the HtmlRAG baselines.

\begin{table}[t]
\centering
\begin{tabular}{lcc}
\toprule
Model & \datasetHotpot & \datasetASQA \\
\midrule

\multicolumn{3}{l}{\emph{Embedding-only retrieval (1000-token citation budget)}} \\
\midrule
\sysBGE                 & 336/1514 (0.222) & 998/1658 (0.602) \\
\sysEmbed               & 479/2225 (0.215) & 1236/2010 (0.615) \\
\sysEmbedNoCtx          & 836/6077 (0.138) & 767/1872 (0.410) \\

\midrule
\multicolumn{3}{l}{\emph{End-to-end pipeline (embedding + contextual filtering)}} \\
\midrule
\sysGenerative          & 538/822 (0.655)  & 1673/2058 (0.813) \\

\bottomrule
\end{tabular}
\caption{Evaluation results across embedding-only retrieval and the end-to-end pipeline.
Each cell reports \#helpful/\#total citations (ratio), evaluated under a fixed 1000-token citation budget.}
\label{tab:eval100-tok}
\end{table}

\subsection{Vector-embeddings-based retrieval}

\paragraph{Baseline system.}
We used the same implementation that was used as a baseline for the HtmlRAG~\cite{10.1145/3696410.3714546}.
\sysBGE consists of the \emph{BGE-Large-EN} embedding model together with the
HTML-side algorithms released by the HtmlRAG authors.
These algorithms are responsible for turning raw HTML pages into embed-able units that respect
the embedding model's input-length constraint (roughly a 500-token limit):
(i) \emph{lossless HTML cleaning} that removes clearly irrelevant content (e.g., scripts/styles) and compresses redundant structure;
(ii) \emph{block-tree construction} that groups the DOM into block-level units at a fixed granularity;
(iii) \emph{embedding-based block scoring} that embeds each block and ranks blocks by similarity to the query; and
(iv) a \emph{budgeted selection} procedure (often described as trim/fill) that retains the highest-scoring blocks while preserving enough surrounding
HTML structure (e.g., required ancestor tags) to keep the pruned result well-formed.
In our experiments, we ported this \sysBGE pipeline verbatim from the HtmlRAG codebase, so the baseline reflects their intended implementation.

\paragraph{Our system.}
Our embedding-based retriever is \sysEmbed, an implementation of the approach we presented in 
Section~\ref{sec:vector-retrieval}.
To isolate algorithmic differences rather than model differences, \sysEmbed uses the same embedding model
as \sysBGE.
Most importantly for interpreting our comparisons, our system \sysEmbed uses the same \emph{BGE-Large-EN} embedding model as \sysBGE;
thus, differences in performance isolate the effect of the retrieval algorithms around the embedding computation
(e.g., how HTML is decomposed, context is preserved, and candidates are assembled), rather than differences in the embedding model itself.
Given that the reported performance of \sysBGE in HtmlRAG is within approximately 5\% of the full HtmlRAG system
across many settings (sometimes slightly better, sometimes slightly worse), suggesting that \sysBGE is a
meaningful proxy baseline for retrieval quality in that work.
For sentence segmentation in \sysEmbed, we use the sentence tokenizer provided by the NLTK toolkit~\cite{bird2009nltk}.

\paragraph{Results.}
Table~\ref{tab:eval100-tok} reports results for embedding-only retrieval under a fixed
budget of 1000 tokens of rendered citation text.
Each entry reports two quantities: \(\#\mathrm{total}\), the total number of citations emitted
by the system within this budget, and \(\#\mathrm{helpful}\), the subset of those citations
judged helpful by the LLM judge.
Because different systems emit citations at different granularities, the value of
\(\#\mathrm{total}\) may vary substantially across systems even though they operate over the
same document collection and under the same token budget.
Accordingly, the ratio \(\#\mathrm{helpful}/\#\mathrm{total}\) should be interpreted as the
primary measure of citation quality, while absolute counts reflect how effectively each system
utilizes the available budget.

Comparing \sysBGE and \sysEmbed, we observe that their helpful-citation ratios are similar across
datasets (e.g., 0.222 vs.\ 0.215 on \datasetHotpot and 0.602 vs.\ 0.615 on \datasetASQA).
However, \sysEmbed produces substantially more citations within the same 1000-token budget
(e.g., 2225 vs.\ 1514 on \datasetHotpot).
This difference arises from citation granularity: \sysEmbed embeds and reports citations at the
level of sentences, while \sysBGE operates at a coarser block level.
Because block-level citations are typically longer, fewer of them fit within the fixed budget.
Taken together, the results indicate that sentence-level retrieval enables \sysEmbed to deliver
more helpful citations under the same budget, while maintaining comparable per-citation quality.

To better understand the impact of global contextualization, we collected results
from an ablation of our system \sysEmbedNoCtx, in which we disable global contextualization.
The ablation results further clarify the role of our design choices.
Disabling global contextualization in \sysEmbedNoCtx yields noticeably lower helpful-citation ratios,
indicating reduced precision.
On \datasetHotpot, \sysEmbedNoCtx nevertheless produces a larger absolute number of helpful
citations; this effect is explained by the removal of global-contextualization material from the budget,
which frees space for additional lower-ranked citations that still contain useful information.

To better understand the impact of using sentences for the unit of embedding, we collected
results from an altered version of our system where instead we take the unit of embedding
to be the \emph{block leaf} element.
Block leaves are the smallest block-level containers (e.g., a \lstinline|<p>|,
\lstinline|<li>|, or \lstinline|<td>|) that still preserve all nested inline markup (links,
emphasis, etc.) but do not contain other block-level regions. This makes them a natural
place to run sentence segmentation without breaking HTML structure.
Because block leaves may be arbitrarily long, they can exceed the input-length limit of the BGE embedding model.
We therefore use the \emph{Qwen~3 0.6B embedding model}~\cite{qwen3-embedding-06b-modelcard} for 
our altered retrieval algorithm, which supports substantially longer
inputs (up to 32k tokens), and was sufficient for all benchmark instances we tested.
Our block-leaf-as-retrieval-unit version performs worse in terms of its ratio, which is $0.176$.
On \datasetHotpot, its total citation count is 1910, similar to the 1514 produced by
\sysBGE, reflecting their shared block-level granularity.

Overall, these results show that sentence-level retrieval combined with global contextualization provides
the best balance between precision and coverage under a fixed citation budget, and that both
coarser block granularity and the removal of global contextualization degrade citation quality in
distinct ways.

\subsection{End-to-end pipeline, with contextual filtering}

We also study our full retrieval pipeline presented in
Section~\ref{sec:generative-retrieval}.
Our contextual filtering step, labeled \sysGenerative, uses an instruction-tuned
language model to explicitly select evidentiary citations from a structured
excerpt produced by embedding-based retrieval.
In our experiments, we use the \emph{Qwen~3 32B} model
\cite{qwen3-techreport,qwen3-32b-modelcard}.
The prompt used to drive our contextual filtering is shown in Appendix~\ref{sec:prompts}.
This prompt enforces a citation-first protocol: the model is instructed to select
only those chunks whose content itself supports the query, to return a minimal
set of such chunks, and to refrain from citing material based on prior knowledge
or latent recall.

Operationally, \sysGenerative builds directly on \sysEmbed.
For each query, we first run the embedding-based retriever and take the prefix of
its ranked results whose combined rendered size fits within a 1000-token budget.
This prefix is passed verbatim to the generative model as a structured excerpt,
with chunk boundaries and identifiers preserved.
Thus, \sysGenerative and \sysEmbed operate over exactly the same candidate material;
the sole difference is the presence of an explicit generative evidence-selection
stage.

\paragraph{Results.}
Table~\ref{tab:eval100-tok} reports results for the end-to-end pipeline, \sysGenerative,
alongside the embedding-only systems under the same 1000-token citation budget.
Across both datasets, \sysGenerative substantially improves helpful-citation ratios.
On \datasetHotpot, \sysGenerative achieves a ratio of 0.655, compared to 0.215 for
\sysEmbed under the same budget, and on \datasetASQA the corresponding ratios are
0.813 and 0.615.
These gains indicate that the contextual filtering is effective at discarding weak or
irrelevant candidates and concentrating the available budget on direct evidence.

We also observe that \sysGenerative changes the effective granularity of what is
counted as a citation.
On \datasetHotpot, \sysGenerative reports fewer citations overall (822 vs.\ 2225 for
\sysEmbed), consistent with the prompt's instruction to return minimal supporting
sets.
On \datasetASQA, the total citation counts are of similar magnitude (2058 vs.\ 2010),
suggesting that our contextual filtering is uncovering additional citations,
which is thanks to the spatial expansion of local contextualization.

Overall, these results show that augmenting embedding-based retrieval with an
explicit contextual-filtering stage yields markedly higher-precision
citation sets while operating on embedding-derived candidates under an identical
citation budget.

\section{Related work}

\paragraph{Structure-aware RAG.}
HtmlRAG~\cite{10.1145/3696410.3714546} and our work share the goal of improving retrieval over HTML documents by leveraging document structure rather than treating HTML as flat text.
However, the two approaches differ substantially in how structure is represented and exploited.
HtmlRAG operates directly over raw HTML source, using a fixed pipeline of preprocessing, block construction, and pruning heuristics to produce embed-able fragments that fit within model input limits.
In contrast, we treat HTML as an internal, tree-structured representation and define retrieval in terms of programmable subdocuments and global-contextualization policies that explicitly capture the structural assumptions under which document fragments are interpretable.
This allows us to decouple embedding and generative models from HTML-specific heuristics, reason formally about context inclusion, and support multiple retrieval strategies over the same structural abstraction.

Our global contextualization concept takes inspiration from a blog post by Lin~\cite{lin2025searchengine}, which argues that naive vector-based retrieval over web documents often fails because many documents are written under conventions that assume surrounding context.
In particular, the post observes that in sources such as manual pages, technical documentation, and reference guides, individual sentences or paragraphs are frequently not self-contained: their meaning depends on nearby headers, section scopes, lists, or other structural cues.
As a result, retrieving an isolated span via embeddings alone can yield excerpts that are syntactically valid but semantically opaque.
Our notion of global contextualization can be viewed as a generalization of this observation.
We formalize global contextualization as 
a document-structural operation that systematically augments retrieved nodes with the minimal surrounding 
context required for interpretability, while remaining compositional.

In this work, we focused on global-contextualization rules for common HTML
structures (e.g., section headers and list spines). A natural next step is to
extend the framework with more domain-specific patterns while preserving the
same abstraction boundary. This raises two concrete questions for future work:
how much of the structure seen in real-world documents can be captured with
such rules, and to what extent can LLMs help infer new rules automatically?

\paragraph{Subdocument.}
The term subdocument has appeared previously in the context of XML processing.
Benedikt and Fundulaki~\cite{10.1007/11601524_9} use the term to describe the result of
\emph{subtree queries}, where an XML document is filtered to produce a smaller document
whose root-to-leaf paths are preserved from the original tree.
Their notion of subdocument is driven by query composition and evaluation in XML databases,
whereas our use of subdocuments is operational: we treat them as programmable,
path-addressed views over document trees that support incremental construction,
cost-bounded context selection, and interaction with LLM and embedding pipelines.

Prior work in hypertext systems has long emphasized fine-grained identity for
document fragments.  The Dexter Hypertext Reference Model~\cite{HalaszSchwartz1994}
provides a canonical abstraction for hypertext systems, distinguishing content,
anchors, and links while deliberately abstracting away from the internal structure
of documents.  Dexter anchors may be understood as naming regions within a document,
but the model does not prescribe how such regions should be rendered, contextualized,
or materialized when detached from their source.  In contrast, our work assumes an
explicit tree-structured document model and focuses on the operational problem that
Dexter leaves open: given an intensional description of a fragment, how to construct
a concrete, well-formed document that preserves the structural assumptions under
which that fragment is interpretable.

A similar emphasis on fine-grained fragment identity appears in Ted Nelson’s
vision for hypertext and transclusion, most fully articulated in
Literary Machines~\cite{Nelson1981}.  Nelson’s work motivates reuse by reference,
provenance preservation, and portion-level addressing, but treats the semantics of
fragment inclusion largely at a conceptual level.  In particular, transclusion
describes what it means to reuse a fragment, but not how to compute the
minimal surrounding structure required for that fragment to stand alone when consumed
outside its original context.  Our notion of subdocuments addresses this gap by
decoupling fragment identity (path sets) from fragment materialization
(pruning after contextualization), yielding a precise extraction semantics that supports
incremental construction, cost-bounded expansion, and reliable use in
retrieval-augmented generation pipelines.

\paragraph{Citation generation.}
LongCite~\cite{zhang2024longcite} and SelfCite~\cite{pmlr-v267-chuang25a} both study the problem of citation generation: 
prompting an LLM with long contexts and training/alignment so that it produces answers with fine-grained, 
sentence-level citations. Our contribution is orthogonal: we focus on retrieval over tree-structured HTML and 
on constructing citations as precise, well-formed, path-addressed document regions. 
The approaches are therefore complementary: LongCite/SelfCite improve how generators cite, 
while our framework improves what is retrieved and how citation regions are represented.

\section{Conclusion}

We presented a retrieval framework that treats document structure as a
first-class concern throughout the retrieval pipeline.
Rather than reducing HTML documents to flat sequences of text spans, our
approach preserves their tree structure and introduces a small set of
abstractions---path sets, subdocuments, and contextualization policies---that
allow retrieval algorithms to reason simultaneously about semantic content and
structural cues.
This design enables the extraction of well-formed excerpts of a desired size
that remain interpretable in isolation, while remaining compatible with both
vector embedding models and generative language models.

Our experimental results demonstrate that these structural choices matter in
practice.
When evaluated under a fixed citation budget and using the same embedding model
as a strong HtmlRAG-derived baseline, our embedding-based retriever achieves
comparable per-citation quality while surfacing substantially more useful
citations.
This effect arises primarily from finer-grained citation units (sentences
rather than blocks) combined with explicit global contextualization, which allows the
system to utilize available budget more effectively.
Ablation studies further confirm that both global contextualization and citation
granularity play critical roles: disabling global contextualization reduces precision,
while reverting to block-level retrieval forfeits the advantages of sentence-level selection.

We also showed that preserving structure enables an effective end-to-end
retrieval pipeline that combines embedding-based search with contextual filtering.
Under the same embedding-derived input and citation budget, the generative
retrieval stage substantially improves helpful-citation ratios on both
benchmarks, indicating that contextual filtering is able to aggressively filter
weak candidates while retaining high-quality evidence.
Importantly, this improvement is achieved without sacrificing provenance:
citations remain tied to precise subdocuments, even though the
contextual filtering reasons over expanded, context-rich views.
This separation between reasoning context and citation identity is central to
maintaining robustness and reproducibility in our full retrieval pipeline.

More broadly, our work suggests that retaining document structure is valuable
not merely for improving chunking heuristics, but for expanding the space of
retrieval operations themselves.
By preserving elements such as sections, tables, lists, and hyperlinks as
addressable structure, retrieval systems can express queries that are difficult
or impossible once documents are flattened, including structure-sensitive
constraints that directly support provenance-aware evidence selection.
We believe these ideas point toward a more principled integration of structured
documents into retrieval-augmented generation pipelines, and offer a foundation
for future work at the intersection of information retrieval, document
understanding, and programming language techniques for structured data.

A practical caveat is that our path-addressed document model is intended as a
high-level specification, not an optimized storage format. In deeply nested
documents, explicit root-to-node paths require space proportional to depth, and
some path operations inherit that cost. In principle, implementations can reduce
this overhead with compact node identifiers and shared-prefix encodings, while
preserving the same retrieval semantics.



\appendix

\section{Prompts used in experiments}
\label{sec:prompts}

\textbf{The following is the prompt we use for scoring citations via our LLM judge.}
{\footnotesize
\begin{verbatim}
You are an expert evaluator who determines whether citations contain useful 
information for answering specific questions.

Your task is straightforward:
1. Read the QUESTION carefully - understand exactly what is being asked
2. Read the CITATION - analyze what information it contains
3. Answer: Does this citation help answer the question? (YES or NO)
4. If YES: Extract the specific part of the citation that helps
5. If YES: Identify which part of the question this citation addresses

EVALUATION PRINCIPLES:

**When to answer YES (helps_answer_question = true):**
- Citation contains facts, names, dates, or details that directly answer 
  the question
- Citation provides information needed to verify or find the answer
- Citation discusses the specific entities, events, or concepts asked about
- Someone reading this citation would get closer to answering the question

**When to answer NO (helps_answer_question = false):**
- Citation only provides generic background without specific answers
- Citation discusses related but different topics
- Citation lacks the specific information the question asks for
- Citation is off-topic or tangentially related

**For helpful_citation_part (if YES):**
- Extract the EXACT text/quote from the citation that helps
- Be specific - quote the relevant sentences or phrases
- Include context if needed but focus on the helpful parts
- If multiple parts help, include all of them

**For answered_question_part (if YES):**
- Extract the EXACT words/phrase from the question text itself that gets answered
- DO NOT write a new sentence - copy the actual text from the question
- Example: If question is "What is the speed of light in a vacuum?", answer: 
  "the speed of light in a vacuum"
- Example: If question is "Who won the Nobel Prize in Physics in 2023?", 
  answer: "Who won the Nobel Prize in Physics in 2023" or just the specific 
  part answered
- If citation answers the full question, return the full question text
- If citation answers only part, return just that portion of the question

**Reasoning:**
- Always provide clear reasoning for your decision
- Explain why the citation does or does not help
- Be specific about what information is present or missing
- Focus on relevance to the actual question asked

**Key Guidelines:**
- Be STRICT: Generic information != helpful answer
- Be PRECISE: Extract exact quotes, not paraphrases
- Be HONEST: If citation doesn't help, say NO
- Be SPECIFIC: Identify exactly what helps and what gets answered
\end{verbatim}
}

\textbf{The prompt listed below is the one we use during local contextualization for the purpose of citation generation.}
{\footnotesize
\begin{verbatim}
You are an evidence selector. Your job is to choose which chunks in the EXCERPT contain evidence
that answers the QUERY. You will see chunks labeled with tags like <chunk3> ... </chunk3>.
Return ONLY a valid JSON array of these OPENING TAGS as strings, copied exactly:
["<chunk3>", "<chunk1>"]
CRITICAL EVIDENCE RULE (anti-hallucination):
- ONLY cite chunks whose *content itself* supports the answer.
- Do NOT cite a chunk just because it triggers knowledge you already have.
- If the excerpt does not contain the needed evidence, return [].
What makes a GOOD citation set:
- A set of chunks which, when read together, contains enough information to answer the query
  fully OR partially.
- A reader should be able to recover the supported part of the answer using ONLY the selected chunks.
What makes a POOR citation set:
- Irrelevant chunks (topic mismatch, background with no claim support).
- Chunks that only contain a title/name/keyword without stating the relevant fact.
- Chunks that merely *hint* at the answer but do not contain the claim, evidence, or reasoning.
- Chunks selected because “the model knows the answer anyway.”
Minimality rule:
- Select the MINIMAL set of chunks that provide evidence.
- Prefer chunks that contain the relevant claim explicitly (names, dates, definitions, numbers, etc.).
If evidence is incomplete:
- Return chunks that provide partial evidence (and omit irrelevant chunks).
- Do NOT fill gaps using prior knowledge; partial evidence is still evidence.
OUTPUT FORMAT RULES:
- Output EXACTLY ONE JSON array of strings.
- Strings must be opening tags exactly as they appear, e.g. "<chunk3>".
- No markdown fences. No extra keys. No commentary before or after the JSON.

Examples (illustrative; do NOT copy content, copy the behavior):
Example A: Complete evidence present
Excerpt contains:
  <chunk2>“The DeLorean was created by John DeLorean.”</chunk2>
Query: “Who created the DeLorean automobile?”
Correct output: ["<chunk2>"]
Example B: Partial evidence present
Excerpt contains:
  <chunk1>“The DeLorean automobile was produced by the DeLorean Motor Company.”</chunk1>
Query: “Who created the DeLorean automobile?”
Correct output: ["<chunk1>"]
(Explanation: This supports only a partial answer about the company; it does NOT name the creator.)
Example C: No evidence in excerpt
Excerpt contains:
  <chunk3># John DeLorean</chunk3>
Query: “Who created the DeLorean automobile?”
Correct output: []
(Reason: a title/name alone does not state the relevant fact; citing it would rely on model knowledge.)
Example D: Model knows the answer but excerpt doesn’t contain it
Query: “What is the capital of France?”
Excerpt contains:
  <chunk0>“France is a country in Europe.”</chunk0>
Correct output: []
(Reason: the excerpt does not contain “Paris”; do not cite based on prior knowledge.)

EXCERPT:
{excerpt_md}
QUERY:
{query}
\end{verbatim}
}

\end{document}